\documentclass[aps,prd,nofootinbib,preprint]{revtex4}

\usepackage{graphicx}
\usepackage{psfrag}

\begin{document}
\def\dirac#1{#1\llap{/}}
\def\pv#1{\mathbf{#1}_\perp}
\newcommand{\ds}{\displaystyle}

\title{$SU_f(3)$-Symmetry Breaking Effects of the $B\to K$
Transition Form Factor in the QCD Light-Cone Sum Rules}
\author{Xing-Gang Wu$^{1}$ \footnote{email: wuxg@cqu.edu.cn},
Tao Huang$^{2}$\footnote{email: huangtao@mail.ihep.ac.cn} and
Zhen-Yun Fang$^{1}$ \footnote{email: zyfang@cqu.edu.cn}}
\address{$^1$Department of Physics, Chongqing University, Chongqing 400044,
P.R. China\\ $^2$Institute of High Energy Physics, Chinese Academy
of Sciences, P.O.Box 918(4), Beijing 100049, P.R. China}

\begin{abstract}
We present an improved calculation of the $B\to K$ transition form
factor with chiral current in the QCD light-cone sum rule (LCSR)
approach. Under the present approach, the most uncertain twist-3
contribution is eliminated. And the contributions from the twist-2
and the twist-4 structures of the kaon wave function are discussed,
including the $SU_f(3)$-breaking effects. One-loop radiative
corrections to the kaonic twist-2 contribution together with the
leading-order twist-4 corrections are studied. The $SU_f(3)$
breaking effect is obtained, $ \frac{F^{B\to
K}_{+}(0)}{F^{B\to\pi}_{+}(0)}=1.16\pm 0.03$. By combining the LCSR
results with the newly obtained perturbative QCD results that have
been calculated up to ${\cal O}(1/m^2_b)$ in Ref.\cite{hwf0}, we
present a consistent analysis of the $B\to K$ transition form factor
in the large and intermediate energy regions. \\


\end{abstract}

\maketitle

\section{Introduction}

There are several approaches to calculate the $B\to {\rm light\;
meson}$ transition form factors, such as the lattice QCD technique,
the QCD light-cone sum rules (LCSRs) and the perturbative QCD (PQCD)
approach. The PQCD calculation is more reliable when the involved
energy scale is hard, i.e. in the large recoil regions; the lattice
QCD results of the $B\to {\rm light \; meson}$ transition form
factors are available only for soft regions; while, the QCD LCSRs
can involve both the hard and the soft contributions below
$m_b^2-2m_b\chi$ ($\chi$ is a typical hadronic scale of roughly
$500$ MeV) and can be extrapolated to higher $q^2$ regions.
Therefore, the results from the PQCD approach, the lattice QCD
approach and the QCD LCSRs are complementary to each other, and by
combining the results from these three methods, one may obtain a
full understanding of the $B\to {\rm light \; meson}$ transition
form factors in its whole physical region. In Refs.\cite{hwbpi,hqw},
we have done a consistent analysis of the $B\to\pi$ transition form
factor in the whole physical region. Similarly, one can obtain a
deep understanding of the $B\to K$ transition form factor in the
physical energy regions by combining the QCD LCSR results with the
PQCD results and by properly taking the $SU_f(3)$ breaking effects
into account.

The $B\to K$ transition form factors are defined as follows:
\begin{eqnarray}
\langle K(p)|\bar{q}\gamma_\mu b|\bar B(p_B)\rangle &=& F_{+}^{B\to
K}(q^2)\left((p+p_B)_{\mu}- \frac{M_B^2-M_K^2}{q^2}q_{\mu}\right) +
F_0^{B\to K}(q^2)\frac{M_B^2-M_K^2}{q^2} q_{\mu} \nonumber\\
&=& 2F_{+}^{B\to K}(q^2) p_{\mu} + F_{-}^{B\to K}(q^2) q_{\mu} \;,
\label{eq:bpi1}
\end{eqnarray}
where the momentum transfer $q=p_B-p$. If we confine ourselves to
discuss the semi-leptonic decays $B\to K l \nu_l$, it is found that
the form factors $F_{-}^{B\to K}(q^2)$ is irrelevant for light
leptons ($l=e,\;\mu$) and only $F_{+}^{B\to K}(q^2)$ matters, i.e.
\begin{equation}
\frac{d\Gamma}{d q^2}(B\to K l \nu_l)=\frac{G_F^2
|V_{tb}V^*_{ts}|^2}{192\pi^3 M_B^3} \lambda^{3/2}(q^2)|F_{+}^{B\to
K}(q^2)|^2 ,
\end{equation}
where $\lambda(q^2)=(M_B^2+M_K^2-q^2)^2-4M_B^2 M_K^2$ is the usual
phase-space factor. So in the following, we shall concentrate our
attention on $F_{+}^{B\to K}(q^2)$.

The $B\to K$ transition form factor has been analyzed by several
groups under the QCD LCSR approach \cite{sumrule,pballsum1,nlosum},
where some extra treatments to the correlation function either from
the B-meson side or from the kaonic side are adopted to improve
their LCSR estimations. It is found that the main uncertainties in
estimation of the $B\to K$ transition form factor come from the
different twist structures of the kaon wave functions. It has been
found that by choosing proper chiral currents in the LCSR approach,
the contributions from the pseudo-scalars' twist-3 structures to the
form factor can be eliminated \cite{huangbpi1,huangbpi2}. In the
present paper, we calculate the $B\to K$ form factor with chiral
current in the LCSR approach to eliminate the most uncertain twist-3
light-cone functions' contributions. And more accurately, we
calculate the ${\cal O}(\alpha_s)$ corrections to the kaonic twist-2
terms. The $SU_f(3)$-breaking effects from the twist-2 and twist-4
kaon wave functions shall also be discussed.

In Ref\cite{hwf0}, we have calculated the $B\to K$ transition form
factor up to ${\cal O}(1/m^2_b)$ in the large recoil region within
the PQCD approach \cite{hwf0}, where the B-meson wave functions
$\Psi_B$ and $\bar\Psi_B$ that include the three-Fock states'
contributions are adopted and the transverse momentum dependence for
both the hard scattering part and the non-perturbative wave
function, the Sudakov effects and the threshold effects are included
to regulate the endpoint singularity and to derive a more reliable
PQCD result. Further more, the contributions from different twist
structures of the kaon wave function, including its
$SU_f(3)$-breaking effects, are discussed. So we shall adopt the
PQCD results of Ref.\cite{hwf0} to do our discussion, i.e. to give a
consistent analysis of the $B\to K$ transition form factor in the
large and intermediate energy regions with the help of the LCSR and
the PQCD results.

The paper is organized as follows. In Sec.II, we present the results
for the $B\to K$ transition form factor within the QCD LCSR
approach. In Sec.III, we discuss the kaonic DAs with $SU_f(3)$
breaking effect being considered. Especially, we construct a model
for the kaonic twist-2 wave function based on the two Gegenbauer
moments $a^K_1$ and $a^K_2$. Numerical results is given in Sec.IV,
where the uncertainties of the LCSR results and a consistent
analysis of the $B\to K$ transition form factor in the large and
intermediate energy regions by combining the QCD LCSR result with
the PQCD result is presented. The final section is reserved for a
summary.

\section{$F_{+}^{B\to K}(q^2)$ in the QCD light-cone sum rule}

The sum rule for $F_{+}^{B\to K}(q^2)$ by including the perturbative
${\cal O}(\alpha_s)$ corrections to the kaonic twist-2 terms can be
schematically written as \cite{huangbpi2,sumrule,sum2}
\begin{equation}
f_{B} F_{+}^{B\to K}(q^2)=\frac{1}{M_B^2}\int_{m_b^2}^{s_0}
e^{(M_B^2-s)/M^2}\left[\rho^{LC}_{T2}(s,q^2)+\rho^{LC}_{T4}(q^2)
\right] ds \;,
\end{equation}
where $\rho^{LC}_{T2}(s,q^2)$ is the contribution from the twist-2
DA and $\rho^{LC}_{T4}(q^2)$ is for twist-4 DA, $f_{B}$ is the
B-meson decay constant. The Borel parameter $M^2$ and the continuum
threshold $s_0$ are determined such that the resulting form factor
does not depend too much on the precise values of these parameters;
in addition the continuum contribution, that is the part of the
dispersive integral from $s_0$ to $\infty$ that has been subtracted
from both sides of the equation, should not be too large, e.g. less
than $30\%$ of the total dispersive integral. The functions
$\rho^{LC}_{T2}(s,q^2)$ and $\rho^{LC}_{T4}(q^2)$ can be obtained by
calculating the following correlation function with chiral current
\begin{eqnarray}
\Pi_{\mu}(p,q) &=& i\int d^4x e^{iq\cdot x}\langle
K(p)|T\{\bar{s}(x)\gamma_\mu(1+\gamma_5)b(x),
\bar{b}(0)i(1+\gamma_5)d(0)\}|0> \nonumber\\
&=& \Pi_+[q^2,(p+q)^2]p_{\mu}+\Pi_-[q^2,(p+q)^2]q_{\mu} \;.
\label{chiral}
\end{eqnarray}
The calculated procedure is the same as that of $B\to\pi$ form
factor that has been done in
Refs.\cite{huangbpi1,huangbpi2,sum2,bagan}. So for simplicity, we
only list the main results for $B\to K$ and highlight the parts that
are different from the case of $B\to\pi$, and the interesting reader
may turn to Refs.\cite{huangbpi2,sum2} for more detailed calculation
technology.

As for $\rho^{LC}_{T2}(s,q^2)$, it can be further written as
\begin{equation}
\rho^{LC}_{T2}(s,q^2)=-\frac{f_K}{\pi}\int_0^1 du \phi_{K}(u,\mu)
{\rm Im}\; T_{T2}\left (\frac{q^2}{m_b^{*2}},
\frac{s}{m_b^{*2}},u,\mu\right ) ,
\end{equation}
where $T_{T2}\left (\frac{q^2}{m_b^{*2}},
\frac{s}{m_b^{*2}},u,\mu\right )$ is the renormalized hard
scattering amplitude, $m_b^*$ stands for the b-quark pole mass
\cite{sum2}. Defining the dimensionless variables
$r_1=q^2/m^{*2}_b$, $r_2=(p+q)^2/m^{*2}_b$ and $\rho=[r_1 +u(r_2
-r_1)-u(1-u)M_K^2/m_b^{*2}]$, up to order $\alpha_s$, we have
\begin{eqnarray}
&&-\frac{{\rm Im}T_{T2}(r_1,r_2,u,\mu)}{\pi} \nonumber\\
&=&\delta(1-\rho)+\frac{\alpha_s(\mu) C_F}{4\pi}\left\{
\delta(1-\rho) \left[ \pi^2 -6 + 3\ln \frac{m_b^{*2}}{\mu^2} -2
\mbox{Li}_2(r_1) +\right.\right. \nonumber\\
&& \left. 2 \mbox{Li}_2(1-r_2)-2\left(\ln\frac{r_2
-1}{1-r_1}\right)^2+2\left(\ln r_2 +\frac{1-r_2}{r_2}
\right)\ln\left(\frac{(r_2-1)^2}{1-r_1}\right)\right]\nonumber\\
&& + \theta(\rho-1)\left[ 8 \left. \frac{\ln(\rho-1)}{\rho-1}
\right|_{+} + 2 \frac{1}{r_2-\rho} \left( \frac{1}{\rho}
-\frac{1}{r_2}\right)\right.\nonumber\\
&& \left.+2 \left( \ln r_2+ \frac{1}{r_2} -2 -2 \ln(r_2-1) + \ln
\frac{m_b^{*2}}{\mu^2} \right) \left. \frac{1}{\rho-1} \right|_{+}
+\frac{1-\rho}{\rho^2} \right. + \nonumber\\
&& \frac{2 (1-r_1)}{(r_1-r_2)(r_2-\rho)} \left( \ln \frac{\rho}
{r_2} -2 \ln \frac{\rho-1}{r_2-1} \right)  -\frac{4 \ln
\rho}{\rho-1} -\nonumber\\
&& \left. \frac{2 (r_2-1)}{(r_1-r_2)(\rho-r_1)}\left( \ln \rho -2
\ln(\rho-1) + 1 - \ln \frac{m_b^{*2}}{\mu^2} \right )\right]
+\nonumber\\
&& \theta(1-\rho) \left[ 2 \left( \ln r_2+ \frac{1}{r_2} -2
\ln(r_2-1) -\ln \frac{m_b^{*2}}{\mu^2} \right) \left.
\frac{1}{\rho-1} \right|_{+} - \right. \nonumber\\
&& \left. \frac{2(1-r_2)}{r_2(r_2-\rho)} \right. \left. \left.
-\frac{2(1-r_1)}{(r_1-r_2)(r_2-\rho)} \left( 1+\ln
\frac{r_2}{(r_2-1)^2}-\ln \frac{m_b^{*2}}{\mu^2} \right) \right]
\right\},
\end{eqnarray}
for the case of $r_1<1$ and $r_2>1$. As for the coefficients of
$\delta(1-\rho)$, the higher power suppressed terms of order ${\cal
O}((M_K^2/m_b^{*2})^2)$ have been neglected due to its smallness.
The dilogarithm function ${\rm
Li}_2(x)=-\int_0^x\frac{dt}{t}\ln(1-t)$ and the operation $``+"$ is
defined by
\begin{equation}
\left. \int d\rho f(\rho)\frac{1}{1-\rho}\right|_{+} =\int d\rho
[f(\rho)-f(1)]\frac{1}{1-\rho}.
\end{equation}
In the calculation, both the ultraviolet and the collinear
divergences are regularized by dimensional regularization and are
renormalized in the $\overline{MS}$ scheme with the totally
anti-commuting $\gamma_5$. And similar to Ref.\cite{sumrule}, to
calculate the renormalized hard scattering amplitude $T_{T2}\left
(\frac{q^2}{m_b^{*2}}, \frac{s}{m_b^{*2}},u,\mu\right )$, the
current mass effects of $s$-quark are not considered due to their
smallness. By setting $M_K\to 0$, it returns to the case of
$B\to\pi$ and it can be found that the coefficients of
$\theta(\rho-1)$ and $\theta(1-\rho)$ agree with those of
Refs.\cite{huangbpi2,sum2}, while the coefficients of
$\delta(1-\rho)$ confirm that of Ref.\cite{sum2} and differ from
that of Ref.\cite{huangbpi2}. The present results can be checked
with the help of the kernel of the Brodsky-Lepage evolution equation
\cite{brodsky}, since the $\mu$-dependences of the hard scattering
amplitude and of the wave function should be compensate to each
other.

As for the sub-leading twist-4 contribution $\rho_{T4}^{LC}(q^2)$,
we calculate it only in the zeroth order in $\alpha_s$, i.e.
\begin{eqnarray}
\frac{\int_{m_b^2}^{s_0}
e^{\frac{M_B^2-s}{M^2}}\rho^{LC}_{T4}(q^2)ds}{M_B^2} &=&
\frac{m_b^{*2} f_K e^{\frac{M_B^2}{M^2}}}{M_B^2}
\left\{\int_{\triangle}^{1} du e^{-\frac{m_b^{*2}-(1-u)(q^2-u M_K^2)
}{u M^2}} \left(\frac{2g_{2}(u)}{uM^2}-
 \frac{8m_b^2[g_1(u)+G_2(u)]}{u^3 M^4}   \right) \nonumber \right. \\
&&+\int_{0}^{1} dv \int D\alpha_i\frac {\theta(\alpha_1+v \alpha_3
-\Delta)}{(\alpha_1+v \alpha_3)^2 M^2} e^{-\frac {m_b^2-(1-\alpha_1-
v \alpha_3) (q^2-(\alpha_1 + v \alpha_3) M_K^2)}{M^2 (\alpha_1+v
\alpha_3)}}\times \nonumber\\
&& \left.\left( 2 \varphi_\perp(\alpha_i)+2
\widetilde\varphi_\perp(\alpha_i)
-\varphi_\parallel(\alpha_i)-\widetilde\varphi_\parallel(\alpha_i)
\right ) \right\} ,
\end{eqnarray}
where $\varphi_\perp(\alpha_i)$,
$\widetilde\varphi_\perp(\alpha_i)$, $\varphi_\parallel(\alpha_i)$
and $\widetilde\varphi_\parallel(\alpha_i)$ are three-particle
twist-4 DAs respectively, and $g_1(u)$ and $g_2(u)$ are two-particle
twist-4 wave functions. Here, $G_2(u)=\int_0^u g_2(v)dv$,
$\triangle=\frac{\sqrt{(s_0-q^2-M_K^2)^2 +4M_K^2(m_b^2-q^2)}
-(s_0-q^2-M_K^2)}{2M_K^2}$ and $s_0$ denotes the subtraction of the
continuum from the spectral integral. By setting $M_K\to 0$ (the
lower integration range of $u$ should be changed to be
$\triangle=\frac{m_b^{*2}-q^2}{s_0-q^2}$ for the case), we return to
the results of $B\to\pi$ \cite{huangbpi2}.

\section{The Distribution amplitudes of kaon}

\subsection{twist-2 DA moments}

Generally, the leading twist-2 DA $\phi_K$ can be expanded as
Gegenbauer polynomials:
\begin{equation}
\phi_K(u,\mu_0) = 6 u (1-u) \left[ 1 + \sum\limits_{n=1}^\infty
a^K_{n}(\mu_0) C_{n}^{3/2}(2u-1)\right]. \label{phiKgen}
\end{equation}
In the literature, only $a^K_{1}(\mu_0)$ is determined with more
confidence level and the higher Gegenbauer moments are still with
large uncertainty and are determined with large errors. Alterative
determinations of Gegenbauer moments rely on the analysis of
experimental data.

The first Gegenbauer moment $a_1^K$ has been studied by the
light-front quark model \cite{quark1}, the LCSR approach
\cite{lcsr1,pballa1k,ballmoments,lenz,zwicky} and the lattice
calculation \cite{lattice1,lattice2} and etc. In Ref.\cite{lcsr1},
the QCD sum rule for the diagonal correlation function of local and
nonlocal axial-vector currents is used, in which the contributions
of condensates up to dimension six and the ${\cal
O}(\alpha_s)$-corrections to the quark-condensate term are taken
into account. The moments derived there are close to that of the
lattice calculation \cite{lattice1,lattice2}, so we shall take
$a^K_1(1{\rm GeV})=0.05\pm 0.02$ to do our discussion. At the scale
$\mu_b=\sqrt{M_B^2-m_b^{*2}}\simeq 2.2$ GeV, $a^K_1(\mu_b)
=0.793a^K_1(1{\rm GeV})$ with the help of the QCD evolution.

The higher Gegenbauer moments, such as $a^K_2$, are still determined
with large uncertainty and are determined with large errors
\cite{sumrule,pballa1k,ballmoments,lcsr1,latt,instat}. For example,
Ref.\cite{instat} shows that the value of $a^K_2$ is very close to
the asymptotic distribution amplitude, i.e. $|a^K_2(1GeV)| \leq
0.04$; while Refs.\cite{pballa1k,lcsr1,latt} gives larger values for
$a^K_2$, i.e. $a^K_2(1GeV)=0.16\pm0.10$ \cite{pballa1k},
$a^K_2(1GeV)=0.27^{+0.37}_{-0.12}$ \cite{lcsr1} and
$a^K_2(2GeV)=0.175\pm 0.065$ \cite{latt}. It should be noted that
the value of $a^K_2$ affects not only the twist-2 structure's
contribution but also the twist-4 structures' contributions, since
the $SU_f(3)$-breaking twist-4 DAs also depend on $a^K_2$ due to the
correlations among the twist-2 and twist-4 DAs as will be shown in
the next subsection. Since the value of $a^K_2$ can not be
definitely known, we take its center value to be a smaller one, i.e.
$a^K_2(1GeV)=0.115$, for easily comparing with the results of
Ref.\cite{sumrule}. Further more, to study the uncertainties caused
by the second Gegenbauer moment $a^K_2$, we shall vary $a^K_2$
within a broader region, e.g. $a^K_2(1GeV)\in [0.05,0.15]$, so as to
see which value is more favorable for $a^K_2$ by comparing with the
PQCD results.

\subsection{Models for the twist-2 and twist-4 DAs}

Before doing the numerical calculation, we need to know the detail
forms for the kaon twist-2 DA and the twist-4 DAs.

As for the twist-2 DA, we do not adopt the Gegenbauer expansion
(\ref{phiKgen}), since its higher Gegenbauer moments are still
determined with large errors whose contributions may not be too
small, i.e. their contributions are comparable to that of the higher
twist structures. For example, by taking a typical value
$a^K_4(1GeV)=-0.015$ \cite{sumrule}, our numerical calculation shows
that its absolute contributions to the form factor is around $1\%$
in the whole allowable energy region, which is comparable to the
twist-4 structures' contributions. Recently, a reasonable
phenomenological model for the kaon wave function has been suggested
in Ref.\cite{hwf0}, which is determined by its first Gegenbauer
moment $a^K_1$, by the constraint over the average value of the
transverse momentum square, $\langle \mathbf{k}_\perp^2
\rangle^{1/2}_K \approx 0.350{\rm GeV}$ \cite{gh}, and by its
overall normalization condition. With the help of such model, a more
reliable PQCD calculation on the $B\to K$ transition form factors up
to ${\cal O}(1/m^2_b)$ have been finished.

In the following, we construct a kaon twist-2 wave function
following the same arguments as that of Ref.\cite{hwf0} but with
slight change to include the second Gegenbauer moment $a^K_2$'s
effect, i.e.
\begin{equation}\label{model}
\Psi_{K}(x,\mathbf{k}_\perp) = [1+B_K C^{3/2}_1(2x-1)+C_K
C^{3/2}_2(2x-1)]\frac{A_K}{x(1-x)} \exp \left[-\beta_K^2
\left(\frac{k_\perp^2+m_q^2}{x}+ \frac{k_\perp^2+m_s^2}
{1-x}\right)\right],
\end{equation}
where $q=u,\; d$, $C^{3/2}_{1,2}(1-2x)$ are the Gegenbauer
polynomial. The constitute quark masses are set to be: $m_q=0.30{\rm
GeV}$ and $m_s=0.45{\rm GeV}$. The four parameters $A_K$, $B_K$,
$C_K$ and $\beta_K$ can be determined by the first two Gegenbauer
moments $a^K_1$ and $a^K_2$, the constraint $\langle
\mathbf{k}_\perp^2 \rangle^{1/2}_K \approx 0.350{\rm GeV}$ \cite{gh}
and the normalization condition $\int^1_0 dx
\int_{k_\perp^2<\mu_0^2} \frac{d^{2}{\bf
k}_{\perp}}{16\pi^3}\Psi_K(x,{\bf k}_{\perp}) =1$. For example, we
have $A_K(\mu_b)=252.044GeV^{-2}$, $B_K(\mu_b)=0.09205$,
$C_K(\mu_b)=0.05250$ and $\beta_K=0.8657GeV^{-1}$ for the case of
$a^K_1(1GeV)=0.05$ and $a^K_2(1GeV)=0.115$. Quantitatively, it can
be found that $B_K$, $C_K$ and $\beta_K$ decreases with the
increment of $a^K_1$; $\beta_K$ decreases with the increment of
$a^K_2$, while $B_K$ and $C_K$ increase with the increment of
$a^K_2$. Under such model, the uncertainty of the twist-2 DA mainly
comes from $a^K_1$ and $a^K_2$. It can be found that the $SU_f(3)$
symmetry is broken by a non-zero $B_K$ and by the mass difference
between the $s$ quark and $u$ (or $d$) quark in the exponential
factor. The $SU_f(3)$ symmetry breaking effect of the leading twist
kaon distribution amplitude has been studied in
Refs.\cite{lcsr1,pballsu} and references therein. The $SU_f(3)$
symmetry breaking in the lepton decays of heavy pseudoscalar mesons
and in the semileptonic decays of mesons have been studied in
Ref.\cite{khlopov}. After doing the integration over the transverse
momentum dependence, we obtain the twist-2 kaon DA,
\begin{eqnarray}
\phi_K(x,\mu_0)&=&\int_{k_\perp^2<\mu_0^2} \frac{d^{2}{\bf
k}_{\perp}}{16\pi^3} \Psi_K(x,{\bf k}_{\perp})\nonumber\\
&=&\frac{A_K}{16\pi^2\beta^2} \left[1+B_K C^{3/2}_1(2x-1)+C_K
C^{3/2}_2(2x-1)\right]\nonumber\\
&&\times\exp \left[-\beta_K^2 \left(\frac{m_q^2}{x}+ \frac{m_s^2}
{1-x}\right)\right]\left[1-\exp\left(-\frac{\beta_K^2\mu^2_0}
{x(1-x)}\right) \right] , \label{phiK}
\end{eqnarray}
where $\mu_0=\mu_b$ for the present case. Then, the Gegenbauer
moments $a^K_n(\mu_0)$ can be defined as
\begin{equation}\label{moments}
a^K_n(\mu_0)=\frac{\int_0^1 dx \phi_K(1-x,\mu_0)C^{3/2}_n(2x-1)}
{\int_0^1 dx 6x(1-x) [C^{3/2}_n(2x-1)]^2} \; ,
\end{equation}
where $\phi_K(1-x,\mu_0)$ other than $\phi_K(x,\mu_0)$ is adopted to
compare the moments with those defined in the literature, e.g.
\cite{lcsr1,pballa1k,ballmoments}, since in these references $x$
stands for the momentum fraction of $s$-quark in the kaon ($\bar
K$), while in the present paper we take $x$ as the momentum fraction
of the light $q$-(anti)quark in the kaon ($K$).

The twist-3 contribution is eliminated by taking proper chiral
currents under the LCSR approach, so we only need to calculate the
subleading twist-4 contributions. The needed four three-particle
twist-4 DAs that are defined in Ref.\cite{braunold} can be expressed
as \cite{pballsum2} \footnote{Similar to Ref.\cite{sumrule}, we
adopt the results that only include the dominant meson-mass
corrections. The less important meson-mass correction terms are not
taken into consideration.}
\begin{eqnarray}
\varphi_{\perp}(\alpha_i) &=& 30 \alpha_3^2(\alpha_2-\alpha_1)\left[
h_{00}+h_{01}\alpha_3+\frac{1}{2}\,h_{10}(5\alpha_3-3)\right] , \nonumber\\
\widetilde{\varphi}_{\perp}(\alpha_i) &=& -30 \alpha_3^2\left[
h_{00}(1-\alpha_3)+h_{01}\Big[\alpha_3(1-\alpha_3)-6\alpha_1\alpha_2\Big]
+h_{10}\Big[\alpha_3(1-\alpha_3)-\frac{3}{2}(\alpha_1^2
+\alpha_2^2)\Big]\right],\nonumber\\
{\varphi}_{\parallel}(\alpha_i) &=& 120 \alpha_1\alpha_2\alpha_3
\left[ a_{10} (\alpha_1-\alpha_2)\right],\nonumber\\
\tilde{\varphi}_{\parallel}(\alpha_i) &=& 120 \alpha_1\alpha_2
\alpha_3 \left[ v_{00} + v_{10} (3\alpha_3-1)\right],
\end{eqnarray}
where
\begin{eqnarray}
h_{00} & = & v_{00} = -\frac{M_K^2}{3}\,\eta_4=-\frac{\delta^2}{3},\nonumber \\
a_{10} & = & \frac{21M_K^2}{8}\eta_4\omega_4 -\frac{9}{20}\, a^K_2
M_K^2=\delta^2\epsilon-\frac{9}{20} a^K_2 M_K^2,\nonumber\\
v_{10} & = & \frac{21M_K^2}{8} \eta_4 \omega_4=\delta^2\epsilon,\nonumber\\
h_{01} & = & \frac{7M_K^2}{4} \eta_4\omega_4 -\frac{3}{20} a^K_2
M_K^2=\frac{2}{3}\delta^2\epsilon-\frac{3}{20} a^K_2 M_K^2 \nonumber
\end{eqnarray}
and
\begin{displaymath}
h_{10} = \frac{7M_K^2}{2} \eta_4\omega_4 + \frac{3}{20} a^K_2
M_K^2=\frac{4}{3}\delta^2\epsilon+\frac{3}{20} a^K_2 M_K^2 ,
\end{displaymath}
with $\eta_4=\delta^2/M_K^2$, $\omega_4=8\epsilon/21$ and
$\delta^2(1GeV)=0.20 GeV^2$ and $\varepsilon(1GeV)=0.53$
\cite{pballsum2}. With the help of QCD evolution, we obtain
$\delta^2(\mu_b)=0.16 GeV^2$ and $\varepsilon(\mu_b)=0.34$. It can
be found that the dominant meson-mass effect are proportional to
$a^K_2$ and $M_K^2$, so if setting $M_K\to 0$ or the value of
$a^K_2$ is quite small, then we return to the results of
Ref.\cite{braunold}. For the remaining two-particle twist-4 wave
functions, their contributions are quite small in comparison to the
leading twist contribution and even to compare with those of the
three-particle twist-4 wave functions. And by taking the leading
meson-mass effect into consideration only, they can be related to
the three-particle twist-4 wave functions through the following way:
\begin{equation}
g_2(u)=\int_0^u d\alpha_1\int_0^{\bar u} d\alpha_2 \frac 1
{\alpha_3} [2\varphi_{\perp}(\alpha_i) -\varphi_{\parallel}
(\alpha_i)]
\end{equation}
and
\begin{equation}
g_1(u)+\int_0^u d v g_2(v)=\frac 1 2 \int_0^u d\alpha_1\int_0^{\bar
u} d\alpha_2 \frac 1 {\alpha_3^2}(\bar u \alpha_1 - u
\alpha_2)[2\varphi_{\perp}(\alpha_i) -
\varphi_{\parallel}(\alpha_i)],
\end{equation}
which lead to
\begin{eqnarray}
g_1(u)  &=&  \frac{\bar{u} u }{6}\left[ -5 \bar{u} u ( 9 h_{00} + 3
h_{01} - 6 h_{10} + 4 \bar{u} h_{01} u + 10 \bar{u} h_{10} u ) +
a_{10} ( 6 +\bar{u} u ( 9 + 80 \bar{u} u ) ) \right]  +\nonumber\\
&& a_{10} {\bar{u}}^3 ( 10-15\bar{u}+6\bar{u}^2)\ln \bar{u} + a_{10}
u^3( 10-15u+6u^2) \ln u , \\
g_2(u)  &=& \frac{5 \bar{u} u ( u -\bar{u})}{2} \left[ 4 h_{00} + 8
a_{10} \bar{u} u - h_{10} (1+5 \bar{u} u ) + 2 h_{01} ( 1 -
\bar{u}u)\right] .
\end{eqnarray}
Similarly, it can be found that when setting $a_2^K\to 0$, the above
expressions of $g_1(u)$ and $g_2(u)$ return to those of
Ref.\cite{braunold}. Here by adopting the relations $
\frac{d\phantom{u}}{du}\, g_2(u) = -\frac{1}{2} \lim_{M_K^2\to 0}
M_K^2 [g_K(u)-\phi_K(u)]$ and $g_1(u)- \int_0^u dv g_2(v) =
\frac{1}{16}\lim_{M_K^2\to 0} M_K^2 {\mathbf A}(u)$, one can
conveniently obtain the higher mass-correction terms for $g_1(u)$
and $g_2(u)$ on the basis of $g_K(u)$ and ${\mathbf A}(u)$ derived
in Refs.\cite{pballsum2,ballmoments}, and numerically, it can be
found that these terms' contributions are indeed small.

\section{Numerical results}

\subsection{basic input}

In the numerical calculations, we use
\begin{eqnarray}
M_B=5.279GeV,\;\; M_K=494MeV,\;\; f_K=160MeV,\;\; f_\pi=131MeV.
\end{eqnarray}

Next, let us choose the input parameters entering into the QCD sum
rule. In general, the value of the continuum threshold $s_0$ might
be different from the phenomenological value of the first radial
excitation mass. Here we set the threshold value of $s_0$ to be
smaller than $s_0^{max}\simeq 34GeV^2$, whose root is slightly
bigger than the mass of the B-meson first radial excitation
predicted by the potential model \cite{potential}. The pole quark
mass $m_b^*$ is taken as $4.7-4.9 GeV$. Another important input is
the decay constant of B meson $f_B$. To keep consistently with the
next-to-leading order calculation of twist-2 contribution, we need
to calculate the two-point sum rule for $f_B$ up to the corrections
of order $\alpha_s$. And in doing the numerical calculation, we
shall adopt the NLO $f_B$ to calculate the NLO twist-2 contribution
and LO $f_B$ for the LO twist-4 contributions for consistence.

\begin{figure}
\centering
\includegraphics[width=0.5\textwidth]{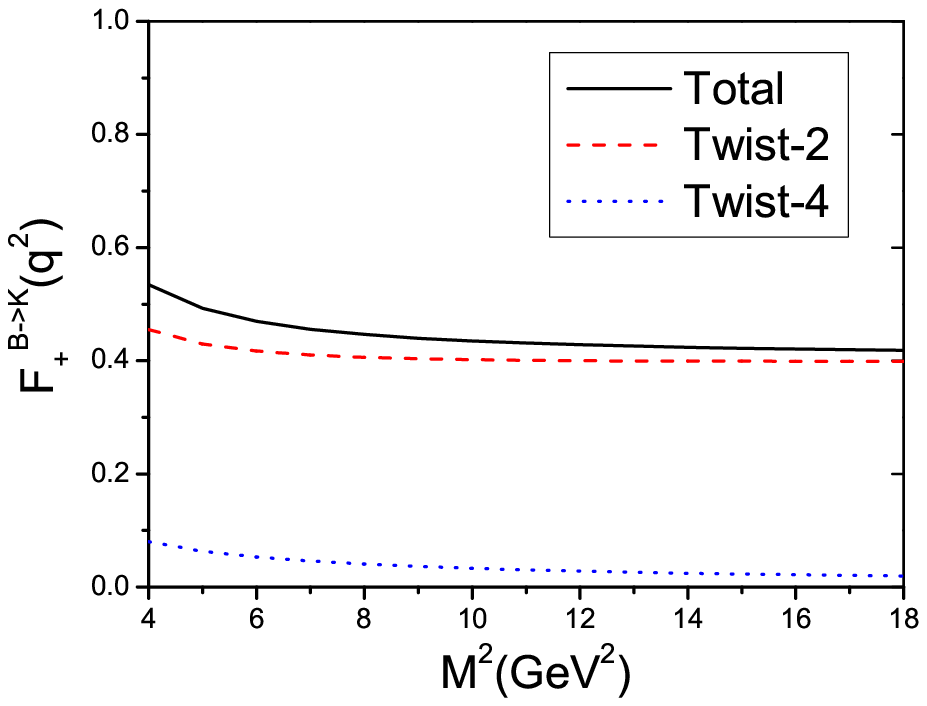}
\caption{$F^{B\to K}_+(q^2)$ as a function of Borel parameter $M^2$
at $q^2=6GeV^2$, where $s_0=33.5GeV^2$, $a_1^K(1GeV)=0.05$,
$a_2^K(1GeV)=0.115$, $m_b^*=4.7GeV$. The solid line stands for the
total contributions, the dashed line is for NLO result of the
twist-2 kaonic wave function and the dotted line is for the LO
result of twist-4 kaonic wave functions.} \label{varyM2}
\end{figure}

The reasonable range for the Borel parameter $M^2$ is determined by
the requirement that the contributions of twist-4 wave functions do
not exceed $10\%$ and those of the continuum states are not too
large, i.e. less than $30\%$ of the total dispersive integration. At
a typical $q^2=6GeV^2$, we draw $F^{B\to K}_+(q^2)$ versus $M^2$ in
Fig.(\ref{varyM2}). It can be found that the contribution from the
kaonic twist-2 wave function slightly increases with the increment
of $M^2$ while the contributions from the kaonic twist-4 wave
functions decreases with the increment of $M^2$, as a result, there
is a platform for $F^{B\to K}_+(q^2)$ as a function of the Borel
parameter $M^2$ for the range of $8GeV^2<M^2<18GeV^2$. For
convenience, we shall always take $M^2=12GeV^2$ to do our following
discussions.

\subsection{uncertainties for the LCSR results}

In the following we discuss the main uncertainties caused by the
present LCSR approach with the chiral current.

The present adopted chiral current approach has a striking advantage
that the twist-3 light-cone functions which are not known as well as
the twist-2 light-cone functions are eliminated, and then it is
supposed to provide results with less uncertainties. In fact, it has
been pointed out that the twist-3 contributions can contribute $\sim
30-40\%$ to the total contribution \cite{bkr} by using the standard
weak current in the correlator, e.g.
\begin{equation}\label{ballsr}
\Pi_{\mu}(p,q) = i\int d^4x e^{iq\cdot x}\langle
K(p)|T\{\bar{s}(x)\gamma_\mu b(x), \bar{b}(0)i\gamma_5 d(0)\}|0> \;.
\end{equation}
If the twist-3 wave functions are not known well, then the
uncertainties shall be large \footnote{A better behaved twist-3 wave
function is helpful to improve the estimations, e.g.
Ref.\cite{piontwist} provides such an example for the pionic case.}.
So in the literature, two ways are adopted to improve the QCD sum
rule estimation on the twist-3 contribution: one is to calculate the
above correlator by including one-loop radiative corrections to the
twist-3 contribution together with the updated twist-3 wave
functions \cite{sumrule}; the other is to introduce proper chiral
current into the correlator, cf. Eq.(\ref{chiral}), so as to
eliminate the twist-3 contribution exactly, which is what we have
adopted. We shall make a comparison of these two approaches in the
following. For such purpose, we adopt the following form for the QCD
sum rule of Ref.\cite{sumrule}, which splits the form factor into
contributions from different Gegenbauer moments:
\begin{equation}\label{sumruleapp}
F^{B\to K}_{+}(q^2)=f^{as}(q^2)+a^K_1(\mu_0)
f^{a^K_1}(q^2)+a^K_2(\mu_0) f^{a^K_2}(q^2)+a^K_4(\mu_0)
f^{a^K_4}(q^2),
\end{equation}
where $f^{as}$ contains the contributions to the form factor from
the asymptotic DA and all higher-twist effects from three-particle
quark-quark-gluon matrix elements, $f^{a^K_1,a^K_2,a^K_4}$ contains
the contribution from the higher Gegenbauer term of DA that is
proportional to $a^K_1$, $a^K_2$ and $a^K_4$ respectively. The
explicit expressions of $f^{as,a^K_1,a^K_2,a^K_4}$ can be found in
Table V and Table IX of Ref.\cite{sumrule}. And in doing the
comparison, we shall take the same DA moments for both methods,
especially the value of $a^K_4(\mu_0)$ is determined from
Eq.(\ref{moments}).

\begin{figure}
\centering
\includegraphics[width=0.5\textwidth]{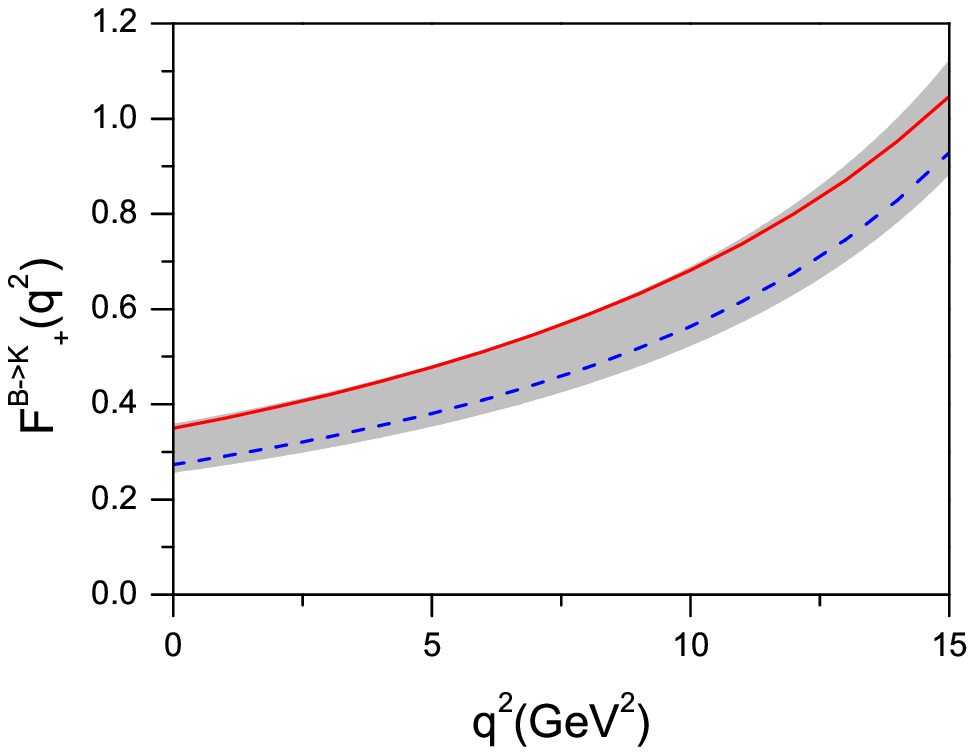}
\caption{$F^{B\to K}_{+}(q^2)$ for $a^K_1(1GeV)\in[0.03,0.07]$,
$a^K_2(1GeV)\in[0.05,0.15]$ and $m_b^* \in [4.7, 4.9]GeV$. The solid
line is obtained with $a^K_1(1GeV)=0.03$, $a^K_2(1GeV)=0.15$ and
$m_b^*=4.9 GeV$; the dashed line is obtained with
$a^K_1(1GeV)=0.07$, $a^K_2(1GeV)=0.05$ and $m_b^*=4.7 GeV$, which
set the upper and the lower ranges of $F^{B\to K}_{+}(q^2)$
respectively. As a comparison, the shaded band shows the result of
Ref.\cite{sumrule} together with its $12\%$ theoretical
uncertainty.} \label{compare}
\end{figure}

We show a comparison of our result of $F^{B\to K}_{+}(q^2)$ with
that of Eq.(\ref{sumruleapp}) in Fig.(\ref{compare}) by varying
$a^K_1(1GeV)\in[0.03,0.07]$, $a^K_2(1GeV)\in[0.05,0.15]$ and
$m_b^*\in [4.7,4.9]GeV$. In Fig.(\ref{compare}) the solid line is
obtained with $a^K_1(1GeV)=0.03$, $a^K_2(1GeV)=0.15$ and $m_b^*=4.9
GeV$; the dashed line is obtained with $a^K_1(1GeV)=0.07$,
$a^K_2(1GeV)=0.05$ and $m_b^*=4.7 GeV$, which set the upper and the
lower ranges of $F^{B\to K}_{+}(q^2)$ respectively. The shaded band
in the figure shows the result of Eq.(\ref{sumruleapp}) within the
same $a^K_1$ and $a^K_2$ region and with its $12\%$ theoretical
uncertainty \cite{sumrule}. It can be found that our present LCSR
results are consistent with those of Ref.\cite{sumrule} within large
energy region $q^2\in[0, 15GeV^2]$. In another words these two
treatments on the most uncertain twist-3 contributions are
equivalent to each other, while the chiral current approach is
simpler due to the elimination of the twist-3 contributions. One may
also observe that in the lower $q^2$ region, different from
Ref.\cite{sumrule} where $F^{B\to K}_{+}(q^2)$ increases with the
increment of both $a^K_1$ and $a^K_2$, the predicted $F^{B\to
K}_{+}(q^2)$ will increase with the increment of $a^K_2$ but with
the decrement of $a^K_1$. This difference is caused by the fact that
we adopt the model wave function (\ref{model}) to do our discussion,
whose parameters are determined by the combined effects of $a^K_1$
and $a^K_2$; while in Ref.\cite{sumrule}, $a^K_1$ and $a^K_2$ are
varied independently and then their contributions are changed
separately.

\begin{table}
\centering
\begin{tabular}{|c||c|c|c||c|c|c|}
\hline ~~~ & \multicolumn{3}{|c||}{LO result} &  \multicolumn{3}{|c|}{NLO result} \\
\hline ~~~ - ~~~ & ~~~$s_0$~~~ & ~~~$M^2$~~~ & ~~~ $f_B$ ~~~&
~~~$s_0$~~~ & ~~~$M^2$~~~ & ~~~ $f_B$ ~~~\\
\hline\hline
~~$m_b=4.7$~~ & 33.5 & 2.80 & 0.165 & 33.5 & 2.80 & 0.219 \\
\hline
$m_b=4.8$ & 33.2 & 2.39 & 0.131 & 33.2 & 2.31 & 0.174 \\
\hline
$m_b=4.9$ & 32.8 & 2.16 & 0.0997 & 32.8 & 2.02 & 0.132 \\
\hline
\end{tabular}
\caption{Parameters for $f_B$, where $m_b$ and $f_B$ are given in
$GeV$, $s_0$ and $M^2$ in $GeV^2$. }\label{tabfb}
\end{table}

Next we discuss the main uncertainties caused by the present LCSR
approach with the chiral current. Firstly, we discuss the
uncertainties of $F^{B\to K}_+(q^2)$ caused by the effective quark
mass $m_b^*$ by fixing $a^K_1(1GeV)=0.05GeV$ and
$a^K_2(1GeV)=0.115GeV$. Under such case, the value of $s_0$, the LO
and NLO vales of $f_B$ should be varied accordingly and be
determined by using the two-point sum rule with the chiral currents,
e.g. to calculate the following two-point correlator:
\begin{equation}
\Pi(q^2)=i\int d^4xe^{iqx}\langle0|\overline{q}(x)(1+\gamma
_5)b(x),\overline{b}(0) (1-\gamma_5)q(0)|0\rangle.
\end{equation}
The sum rule for $f_B$ up to NLO can be obtained from
Ref.\cite{sumrulefb} through a proper combination of the scalar and
pseudo-scalar results shown there \footnote{One needs to change the
$c$-quark mass to the present case of $b$-quark mass and we take
$\langle\frac{\alpha_s}{\pi}G^a_{\mu\nu}G^{a
{\mu\nu}}\rangle=2\times(0.33GeV)^4$ \cite{con1} and
$\alpha_s\langle q\bar{q}\rangle^2=0.162\times10^{-3}GeV^6$
\cite{sumrulefb} to do the numerical calculation.}, which can be
schematically written as
\begin{equation} \label{sumfb}
f_B^2 M_B^2 e^{-M^2_B/M^2}=\int_{m_b^2}^{s_0}\rho^{tot}(s)
e^{-s/M^2}ds ,
\end{equation}
where the spectral density $\rho^{tot}(s)$ can be read from
Ref.\cite{sumrulefb}. The Borel parameter $M^2$ and the continuum
threshold $s_0$ are determined such that the resulting form factor
does not depend too much on the precise values of these parameters;
in addition, 1) the continuum contribution, that is the part of the
dispersive integral from $s_0$ to $\infty$, should not be too large,
e.g. less than $30\%$ of the total dispersive integral; 2) the
contributions from the dimension-six condensate terms shall not
exceed $15\%$ for $f_B$. Further more, we adopt an extra criteria as
suggested in Ref.\cite{sumrule} to derive $f_B$: i.e. the derivative
of the logarithm of Eq.(\ref{sumfb}) with respect to $1/M^2$ gives
the B-meson mass $M_B$,
\begin{displaymath}
M_B^2=\int_{m_b^2}^{s_0}\rho^{tot}(s) e^{-s/M^2} s ds {\Bigg /}
\int_{m_b^2}^{s_0}\rho^{tot}(s) e^{-s/M^2}ds ,
\end{displaymath}
and we require its value to be full-filled with high accuracy $\sim
0.1\%$. These criteria define a set of parameters for each value of
$m^*_b$. Some typical values of $f_B$ are shown in TAB.\ref{tabfb},
where $f_B$ is taken as the extremum within the reasonable region of
$(M^2,s_0)$ and the value of $m_b^*$ is taken as \cite{mbmass}:
$m_b^*\simeq 4.8 \pm 0.1GeV$ . $f_B$ decreases with the increment of
$m^*_b$. The NLO result agrees with the first direct measurement of
this quantity by Belle experiment $f_B=229^{+36}_{-31}({\rm stat})
^{+34}_{-37}({\rm syst})$ MeV from the measurement of the decay
$B^{-}\to\tau\bar\nu_\tau$ \cite{belle}.

\begin{figure}
\centering
\includegraphics[width=0.5\textwidth]{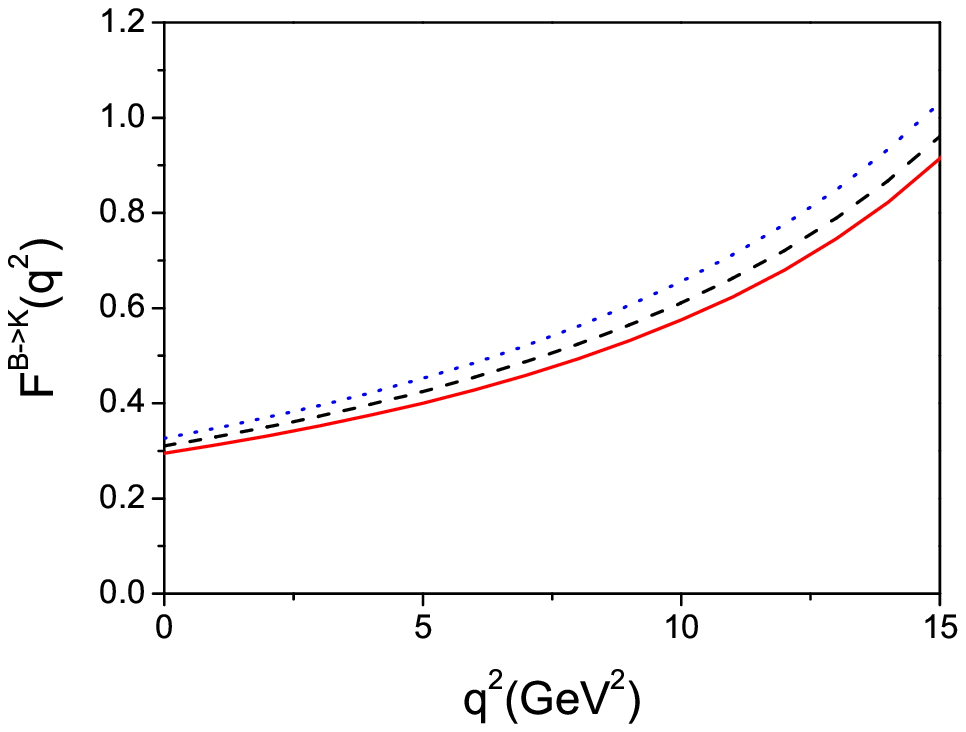}
\caption{$F^{B\to K}_+(q^2)$ as a function of $q^2$ with varying
$m_b^*$. The solid, dashed line and the dash-dot line are for
$m_b^*=4.7GeV$, $4.8GeV$ and $4.9GeV$ respectively, where
$a^K_1(1GeV)=0.05GeV$ and $a^K_2(1GeV)=0.115GeV$.} \label{varymb}
\end{figure}

The value of $F^{B\to K}_+(q^2)$ for three typical values of
$m_b^*$, i.e. $m_b^*=4.7GeV$, $4.8GeV$ and $4.9GeV$ respectively,
are shown in Fig.(\ref{varymb}). $F^{B\to K}_+(q^2)$ increases with
the increment of $m^*_b$. It can be found that the uncertainty of
the form factor caused by $m_b^*\in [0.47GeV,0.49GeV]$ is $\sim 5\%$
at $q^2=0$ and increases to $\sim 9\%$ at $q^2=14GeV^2$. By taking a
more accurate $m_b^*$, e.g. $m_b^*=(4.80\pm0.05)GeV$ as suggested by
Ref.\cite{sumrule}, the uncertainties can be reduced to $\sim 3\%$
at $q^2=0$ and $\sim 5\%$ at $q^2=14GeV^2$.

\begin{figure}
\centering
\includegraphics[width=0.5\textwidth]{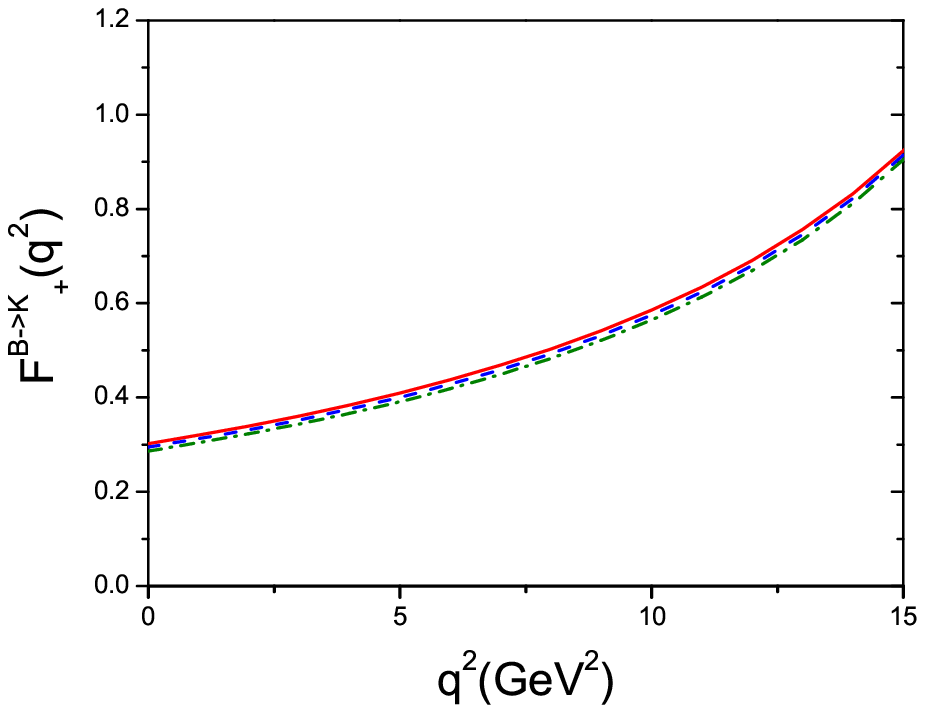}
\caption{$F^{B\to K}_+(q^2)$ as a function of $q^2$ with varying
$a^K_1(1GeV)$, where $a^K_2(1GeV)=0.115$. The solid line, the dashed
line and the dash-dot line are for $a^K_1(1GeV)=0.03$, $0.05$ and
$0.07$ respectively. } \label{varya1k}
\end{figure}

\begin{figure}
\centering
\includegraphics[width=0.5\textwidth]{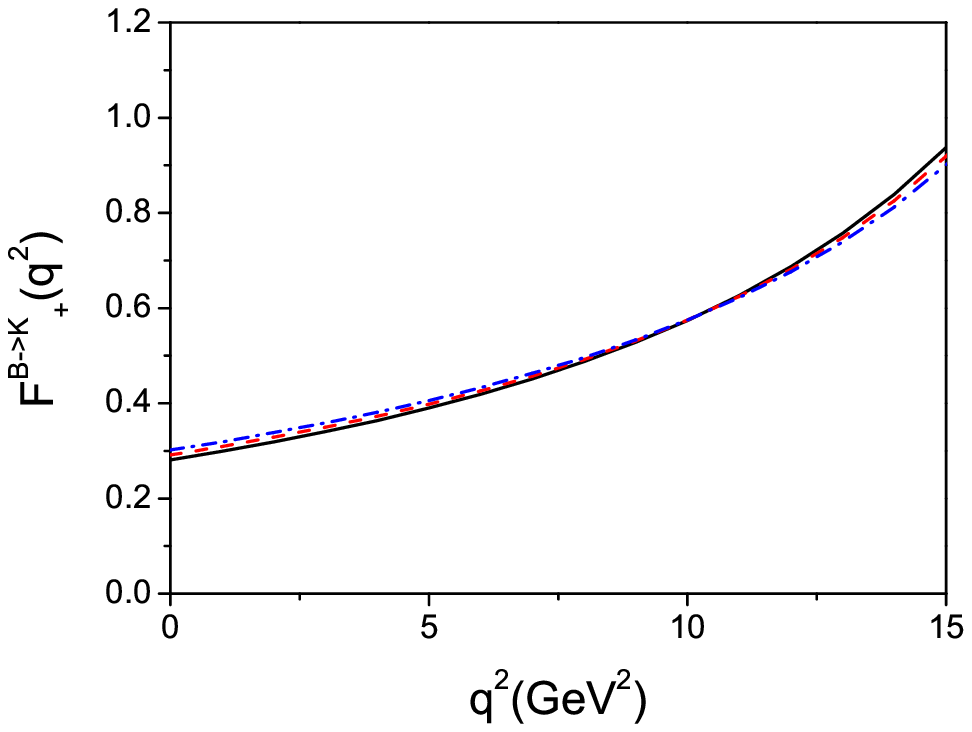}
\caption{$F^{B\to K}_+(q^2)$ as a function of $q^2$ with varying
$a^K_2(1GeV)$, where $a^K_1(1GeV)=0.05$. The solid line, the dashed
line and the dash-dot line are for $a^K_2(1GeV)=0.05$, $0.10$ and
$0.15$ respectively. } \label{varya2k}
\end{figure}

Secondly, we discuss the uncertainties of $F^{B\to K}_+(q^2)$ caused
by the twist-2 wave function $\Psi_K$, i.e. the two Gegenbauer
moments $a^K_1(1GeV)$ and $a^K_2(1GeV)$. For such purpose, we fix
$s_0=33.5GeV^2$ and $m_b^*=4.7GeV$. To discuss the uncertainties
caused by $a^K_1(1GeV)$, we take $a^K_2(1GeV)=0.115$. $F^{B\to
K}_+(q^2)$ for three typical $a^K_1(1GeV)$, i.e. $a^K_1(1GeV)=0.03$,
$0.05$ and $0.07$ respectively, are shown in Fig.(\ref{varya1k}).
$F^{B\to K}_+(q^2)$ decreases with the increment of $a^K_1$. It can
be found that the uncertainty of form factor caused by $a^K_1(1{\rm
GeV})\in [0.03,0.07]$ is small, i.e. it is about $3\%$ at $q^2=0$
and becomes even smaller for larger $q^2$. Similarly, to discuss the
uncertainties caused by $a^K_2(1GeV)$, we fix $a^K_1(1{\rm
GeV})=0.05$. Since the value of $a^K_2$ is less certain than
$a^K_1$, so we take three typical values of $a^K_2(1GeV)$ with
broader separation to calculate $F^{B\to K}_+(q^2)$, i.e.
$a^K_1(1GeV)=0.05$, $0.10$ and $0.15$ respectively. The results are
shown in Fig.(\ref{varya2k}). It can be found that the uncertainty
of the form factor caused by $a^K_2(1{\rm GeV})\in [0.05,0.15]$ is
also small, i.e. it is about $5\%$ at $q^2=0$ and becomes smaller
for larger $q^2$. $F^{B\to K}_+(q^2)$ increases with the increment
of $a^K_2$ in the lower energy region $q^2<10GeV^2$ and decreases
with the increment of $a^K_2$ in the higher energy region $q^2 >
10GeV^2$.

As a summary, a more accurate values for $m_b^*$, $a^K_1$ and
$a^K_2$ shall be helpful to derive a more accurate result for the
form factor. Our results favor a smaller $a^K_2$ to compare with the
form factor in the literature, e.g. $a^K_2(1GeV)\leq 0.15$. And
under such region, the uncertainties from $a^K_2$ is small, i.e. its
uncertainty is less than $5\%$ for $a^K_2(1GeV)\in[0.05, 0.15]$. It
can be found that by varying $a^K_1(1GeV)\in[0.03,0.07]$ and
$a^K_2(1GeV)\in[0.05,0.15]$, the kaonic twist-4 wave functions'
contribution is about $6\%$ of the total contribution at $q^2=0$.
The uncertainties of $a^K_1$ shows that the $SU_f(3)$-breaking
effect is small but it is comparable to that of the higher twist
structures' contribution. So the $SU_f(3)$ breaking effect and the
higher twist's contributions should be treated on the equal footing.
Using the chiral current in the correlator, as shown in
Eq.(\ref{chiral}), the theoretical uncertainty can be remarkably
reduced. And our present LCSR results are consistent with those of
Ref.\cite{sumrule} within large energy region $q^2\in[0, 15GeV^2]$,
which is calculated with the correlator (\ref{ballsr}) and includes
one-loop radiative corrections to twist-2 and twist-3 contributions
together with the updated twist-3 wave functions. In another words
these two approaches are equivalent to each other in some sense,
while the chiral current approach is simpler due to the elimination
of the more or less uncertain twist-3 contributions. For higher
energy region $q^2>15GeV^2$, the LCSR approach is no longer
reliable. Therefore the lattice calculations, would be extremely
useful to derive a more reliable estimation on the high energy
behaviors of the form factors.

\subsection{$SU_f(3)$ breaking effect of the form factor within the LCSR}

\begin{figure}
\centering
\includegraphics[width=0.5\textwidth]{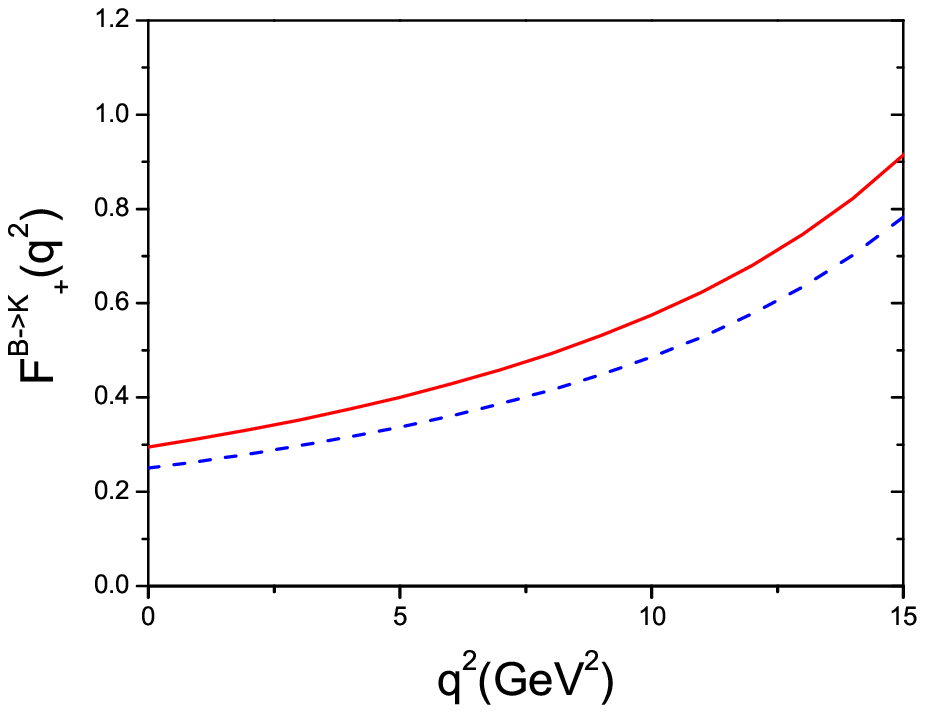}
\caption{Comparison of $F^{B\to K}_+(q^2)$ and $F^{B\to
\pi}_+(q^2)$, where $m_b^*=4.7GeV$, $s_0=33.5GeV^2$,
$f^{LO}_B=0.165GeV$, $f^{NLO}_B=0.219GeV$, $a^K_1(1GeV)=0.05$ and
$a^{\pi/K}_2(1GeV)=0.115$. The solid line and the dashed line are
for $F^{B\to K}_+(q^2)$ and $F^{B\to \pi}_+(q^2)$ respectively. }
\label{pik}
\end{figure}

To have an overall estimation of the $SU_f(3)$ breaking effect, we
make a comparison of the $B\to\pi$ and $B\to K$ form factors:
$F^{B\to\pi}_{+}(q^2)$ and $F^{B\to K}_{+}(q^2)$. The formulae for
$F^{B\to\pi}_{+}(q^2)$ can be conveniently obtained from that of
$F^{B\to K}_{+}(q^2)$ by taking the limit $M_K\to 0$. In doing the
calculation for $F^{B\to\pi}_{+}(q^2)$, we directly use the
Gegenbauer expansion for pion twist-2 DA, because different to the
kaonic case, now the higher Gegenbauer terms' contributions are
quite small even in comparison to the twist-4 contributions, e.g. by
taking $a^\pi_4(1GeV)=-0.015$ \cite{sumrule}, our numerical
calculation shows that its absolute contributions to the form factor
is less than $0.5\%$ in the whole allowable energy region. We show a
comparison of $F^{B\to K}_+(q^2)$ and $F^{B\to \pi}_+(q^2)$ in
Fig.(\ref{pik}) with the parameters taken to be $m_b^*=4.7GeV$,
$s_0=33.5GeV^2$, $f^{LO}_B=0.165GeV$, $f^{NLO}_B=0.219GeV$,
$a^K_1(1GeV)=0.05$ and $a^{\pi/K}_2(1GeV)=0.115$. Secondly, by
varying $m_b^* \in [4.7, 4.9]GeV$, $a^K_1(1GeV)\in [0.03, 0.07]$ and
$a^{\pi/K}_2(1GeV) \in [0.05,0.15]$, we obtain
$F^{B\to\pi}_{+}(0)\in[0.239,0.294]$ and $F^{B\to
K}_{+}(0)\in[0.273,0.349]$. Then we obtain $ \frac{F^{B\to
K}_{+}(0)}{F^{B\to\pi}_{+}(0)}=1.16\pm 0.03$, which favors a small
$SU_f(3)$ breaking effect and is consistent with the PQCD estimation
$1.13\pm0.02$ \cite{hwf0}, the QCD sum rule estimations, e.g.
$[F^{B\to K}_{+}(0)/F^{B\to\pi}_{+}(0)]\approx 1.16$
\cite{sumrule}\footnote{To estimate the ratio $[F^{B\to
K}_{+}(0)/F^{B\to\pi}_{+}(0)]$ from Ref.\cite{sumrule}, we take
$a^K_1(1GeV)= 0.05\pm 0.02$.}, $1.08^{+0.19}_{-0.17}$ \cite{khod}
and $1.36^{+0.12}_{-0.09}$ \cite{lcsr1} respectively, and a recently
relativistic treatment that is based on the study of the
Dyson-Schwinger equations in QCD, i.e. $[F^{B\to
K}_{+}(0)/F^{B\to\pi}_{+}(0)]=1.23$ \cite{roberts}.

\subsection{consistent analysis of the form factor within
the large and the intermediate energy regions}

Recently, Ref\cite{hwf0} gives a calculation of the $B\to K$
transition form factor up to ${\cal O}(1/m^2_b)$ in the large recoil
region within the PQCD approach \cite{hwf0}, where the B-meson wave
functions $\Psi_B$ and $\bar\Psi_B$ that include the three-Fock
states' contributions are adopted and the transverse momentum
dependence for both the hard scattering part and the
non-perturbative wave function, the Sudakov effects and the
threshold effects are included to regulate the endpoint singularity
and to derive a more reliable PQCD result. Further more, the
uncertainties for the PQCD calculation of the $B\to K$ transition
form factor has been carefully studied in Ref.\cite{hwf0}. So we
shall adopt the PQCD results of Ref.\cite{hwf0} to do our
discussion. Only we need to change the twist-2 kaon wave function
$\Psi_K$ used there to the present one as shown in Eq.(\ref{model}).

\begin{figure}
\centering
\includegraphics[width=0.5\textwidth]{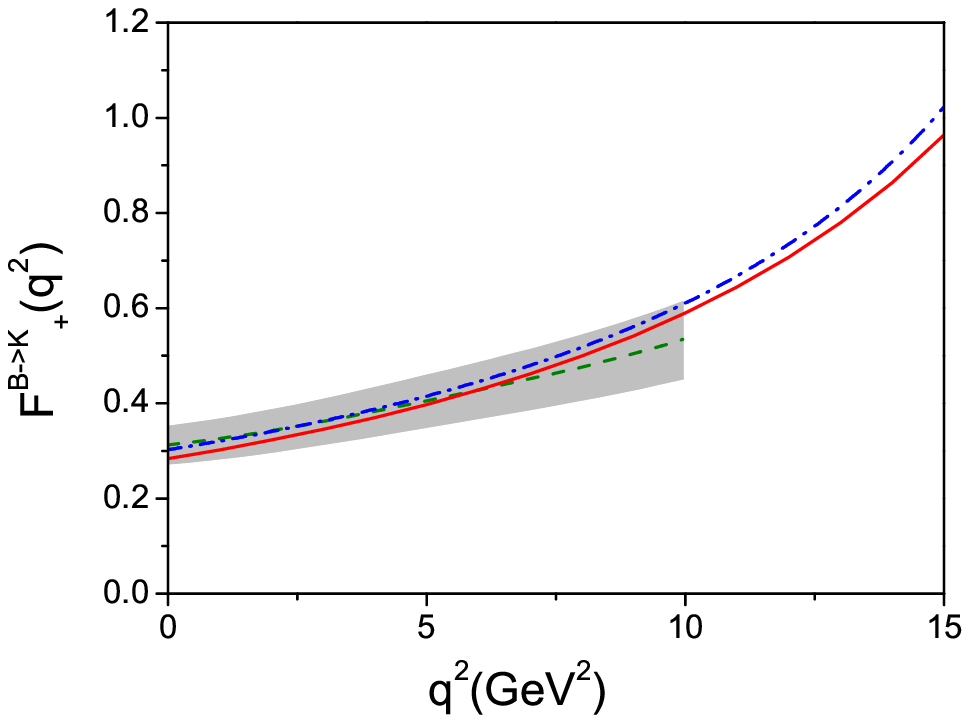}
\caption{LCSR and PQCD results for $F^{B\to K}_+(q^2)$. The solid
line is for our LCSR result, the dash-dot line is for the LCSR
result of Ref.\cite{sumrule} with $a^K_1(1GeV)=0.07$ and
$a^K_2(1GeV)=0.05$. The shaded band is the PQCD result with
$\bar\Lambda\in[0.50,0.55]$ and $\delta\in[0.25,0.30]$, where the
dashed line is for the center values $\bar\Lambda=0.525$ and
$\delta=0.275$, the upper edge of the band is for $\bar\Lambda=0.50$
and $\delta=0.30$ and the lower edge of the band is for
$\bar\Lambda=0.55$ and $\delta=0.25$. } \label{lcsrpqcd}
\end{figure}

We show the LCSR results together with the PQCD results in
Fig.(\ref{lcsrpqcd}). In drawing the figure, we take
$a^K_1(1GeV)=0.07$, $a^K_2(1GeV)=0.05$ and $m_b^*=4.8GeV$. And the
uncertainties of these parameters cause about $\sim 10\%$ errors for
the LCSR calculation. While for the PQCD results, we should also
consider the uncertainties from the B-meson wave functions, i.e. the
values of the two typical parameters $\bar\Lambda$ and $\delta$, and
we take $\bar\Lambda\in[0.50,0.55]$ and $\delta\in[0.25,0.30]$
\cite{hwf0}. It can be found that the PQCD results can match with
the LCSR results for small $q^2$ region, e.g. $q^2<10GeV^2$. Then by
combining the PQCD results with the LCSR results, we can obtain a
consistent analysis of the form factor within the large and the
intermediate energy regions. Inversely, if the PQCD approach must be
consistent with the LCSR approach, then we can obtain some
constraints to the undetermined parameters within both approaches.
For example, according to the QCD LCSR calculation, the form factor
$F^{B\to K}_+(q^2)$ increases with the increment of b-quark mass,
then the value of $m_b$ can not be too large or too small \footnote{
Another restriction on $m_b$ is from the experimental value
\cite{belle} on $f_B$.}, i.e. if allowing the discrepancy between
the LCSR result and the PQCD results to be less than $15\%$, then
$m_b^*$ should be around the value of $4.8\pm 0.1GeV$.

\section{Summary}

In the paper, we have calculated the $B\to K$ transition form factor
by using the chiral current approach under the LCSR framework, where
the $SU_f(3)$ breaking effects have been considered and the twist-2
contribution is calculated up to next-to-leading order. It is found
that our present LCSR results are consistent with those of
Ref.\cite{sumrule} within large energy region $q^2\in[0, 15GeV^2]$,
which is calculated with the conventional correlator (\ref{ballsr})
and includes one-loop radiative corrections to twist-2 and twist-3
contributions together with the updated twist-3 wave functions. And
our present adopted LCSR approach with the chiral current is simpler
due to the elimination of the more or less uncertain twist-3
contributions.

The uncertainties of the LCSR approach have been discussed,
especially we have found that the second Gegenbauer moment $a^K_2$
prefers asymptotic-like smaller values. By varying the parameters
within the reasonable regions: $m_b^* \in [4.7, 4.9]GeV$,
$a^K_1(1GeV)\in [0.03, 0.07]$ and $a^{\pi/K}_2(1GeV) \in
[0.05,0.15]$, we obtain $F^{B\to\pi}_{+}(0)=0.267\pm 0.026$ and
$F^{B\to K}_{+}(0)=0.311\pm0.038$, which are consistent with the
PQCD and the QCD sum rule estimations in the literature.
Consequently, we obtain $ \frac{F^{B\to
K}_{+}(0)}{F^{B\to\pi}_{+}(0)}=1.16\pm 0.03$, which favors a small
$SU_f(3)$ breaking effect. Also, it has been shown that one can do a
consistent analysis of the $B\to K$ transition form factor in the
large and intermediate energy regions by combining the QCD LCSR
result with the PQCD result. The PQCD approach can be applied to
calculate the $B\to K$ transition form factor in the large recoil
regions; while the QCD LCSR can be applied to intermediate energy
regions. Combining the PQCD results with the QCD LCSR, we can give a
reasonable explanation for the form factor in the low and
intermediate energy regions. Further more, the lattice estimation
shall help to understand the form factors' behaviors in even higher
momentum transfer regions, e.g. $q^2>15GeV^2$. So, we suggest such a
lattice calculation can be helpful. Then by comparing the results of
these three approaches, the $B\to K$ transition form factor can be
determined in the whole kinematic regions.

\begin{center}
\section*{Acknowledgements}
\end{center}

This work was supported in part by the Natural Science Foundation of
China (NSFC) and by the Grant from Chongqing University. This work
was also partly supported by the National Basic Research Programme
of China under Grant NO. 2003CB716300. The authors would like to
thank Z.H. Li, Z.G. Wang and F.Zuo for helpful discussions on the
determination of $f_B$.

\end{document}